# The Rosetta mission orbiter Science overview – the comet phase

M.G.G.T. Taylor (1), N. Altobelli (2), B. J. Buratti (3) and M. Choukroun (3)

*(1) ESA/ESTEC, Noordwijk, Netherlands (2) ESA/ESAC, Villanueva de al Canada, Spain (3) JPL/California Institute of Technology, Pasadena, USA*

## Summary

The International Rosetta Mission was launched in 2004 and consists of the orbiter spacecraft Rosetta and the lander Philae. The aim of the mission is to map the comet 67-P/Churyumov-Gerasimenko by remote sensing, to examine its environment insitu and its evolution in the inner solar system. Rosetta was the first spacecraft to rendezvous and orbit a comet, accompanying it as it passes through the inner solar system, and to deploy a lander, Philae and perform in-situ science on the comet surface. The primary goals of the mission were to: characterize the comet's nucleus; examine the chemical, mineralogical and isotopic composition of volatiles and refractories; examine the physical properties and interrelation of volatiles and refractories in a cometary nucleus; study the development of cometary activity and the processes in the surface layer of the nucleus and in the coma; detail the origin of comets, the relationship between cometary and interstellar material and the implications for the origin of the solar system; characterize asteroids, 2867 Steins and 21 Lutetia. This paper presents a summary of mission operations and science, focusing on the Rosetta orbiter component of the mission during its comet phase, from early 2014 up to September 2016.

## Introduction

The Rosetta Mission is the third cornerstone mission the ESA programme Horizon 2000 **[1][2],** with a prime aim to chase down and escort a comet as it passed through the inner solar system, as well as characterizing two asteroids on the way. Small bodies such as asteroids and comets are considered the left over material from the formation of the planets and the solar system. Comets are of particular interest as they have spent most of their life located very far from the Sun, thus retaining the most primordial elements of the solar system and providing an insight into the composition of the interstellar cloud preceding our Sun and planets. The names Rosetta and Philae were taken from the Rosetta Stone **[3]** and the Philae obelisk **[4]** (one of two found at Philae in upper Egypt). These Egyptian artifacts have inscriptions in different texts and were key in deciphering the Egyptian hieroglyphs. Naming the spacecraft after these ancient Egyptian artifacts seemed highly appropriate, as the Rosetta and Philae spacecraft will play a key role in unraveling the "language" of the ancient solar system.
The mission was originally targeted to visit comet 46P/Wirtanen **[5]** but due to launcher issues and subsequent launch delay, in 2003 comet 67P/ Churyumov-Gerasimenko was selected as the new target. Comet 67P/ Churyumov-Gerasimenko was discovered by Klim Ivanovich Churyumov and Svetlana Ivanova Gerasimenko in September 1969. Highly focused campaigns, utilizing professional and amateur teams, were carried out to characterize the comet **[5][6]**, combining a number of ground-based and near –Earth telescopes, from which its period, 6.55 years, perihelion 1.24 AU and aphelion 5.68 AU were refined as well as its nucleus shape, ranging from 4.4 - 5.1 km depending on spin direction **[5]** with a rotation period of 12.76 hours **[6]**.

*Author for correspondence (mtaylor@esa.int).
†Present address: Scientific Support Office, Directorate of Science,
ESTEC, European Space Agency, Keplerlaan 1,
Noordwijk, 2201AZ, The Netherlands

Analysis of the comets activity was made using heliocentric light curves **[7]** providing valuable input for planned observations at the comet, once Rosetta arrived.
Following its launch in March 2004, the Rosetta mission underwent three Earth flybys and one Mars flyby to achieve the correct trajectory to capture the comet, including flybys of asteroid on 2867 Steins **[8]** and 21 Lutetia **[9]**. From June 2011- January 2014 the spacecraft passed through a period of hibernation, due to lack of available power for full payload operation. Following successful hibernation exit and subsequent instrument commissioning, it successfully rendezvoused with the comet in August 2014. Following an intense period of mapping and nucleus characterisation, a landing site for Philae was selected, and on 12 November 2014, Philae was successfully deployed. Rosetta then embarked on the main phase of the mission, observing the comet on its way into and away from perihelion in August 2015. At the time of writing the mission is due to terminate with the Rosetta orbiter impacting the comet surface on 30 September 2016. The evolution of the spacecraft comet- distance during the mission comet phase is shown in figure 1, along with a more detailed description of the spacecraft trajectory with respect to the comet from January 2015-August 2016 in Figure 2.

This paper is derived from an overview presentation of the Rosetta orbiter component of the mission at a special discussion meeting held at the Royal Society in London from 14-15 June 2016, entitled "Cometary Science after Rosetta", where the ongoing science of the Rosetta mission was put into context of cometary science as a whole. Its aim is to provide a glimpse at the initial insight Rosetta has provided us.

## Mission Science

The prime scientific goals of the Rosetta mission are shown in **Table 1**. These goals are outlined in the Rosetta Science Management Plan **[1]**, with all but the final objective focused on the comet phase. These high level goals formed the basis of the master science plan (MSP) of the mission which consisted of broader themes and more specific items, including any evolution of the underlying science as a result of advances in cometary science **[10]** during the Rosetta cruise phase. The MSP subsequently fed into the science planning component **[11] [12]** of the mission operations **[13]**. At different periods of the mission, different goals have been prioritized, with an aim to cover them all at a sufficient level at all heliocentric distances. Overall, these can be broken down to 5 phases: Pre-landing (from ~ 4.4 to 2.95 AU, March/April to November 2014), "first times" (from 2.95 to 2.1 AU, November 2014-March 2015), development of cometary activity (2.1 to 1.25 AU , March 2015- August 2015, perihelion), comparison with pre-perihelion (1.25 – 2.01 AU, August to December 2015) and extension (2.01 – 3.8 AU, January -September 2016). The extension phase provides a larger heliocentric range of observations, to compare with pre-landing, along with a number specific activities focusing on near comet observations and an excursion into the night-side of the comet. At the time of writing, operationally the spacecraft is in the extension phase, but the science activity has only addressed the first few phases. The mission is due to end on 30 September 2016 by impacting the surface of the comet. This end of mission scenario maximizes the science possible given power and data rate constraints and provides the unique opportunity to access < 5 km altitudes, not obtained in the mission so far.

## Pre-landing and first times

Following hibernation exit in January 2014, the spacecraft was some 9 million kilometres from the comet, at around 4.49 AU from the Sun. Instrument commissioning began and the comet nucleus was gradually resolved, with observations showing a change in sidereal rotation period since its previous apparition by around ~ 1300 seconds to P=12.40 hours **[14]** and evidence of activity already in early 2014 **[15][16]** at around 4.3 AU **[17]**, with activity close to expected levels **[18]**. This change in period has been associated with the influence of the sublimation activity of the irregular shape of the nucleus **[19]**. Following comet rendezvous in August 2014, the Rosetta spacecraft crossed to within 100 km of the bi-lobed, "duck" like nucleus, eventually reaching 10 km in October prior to lander deployment. This period was dominated by the characterization of the comet nucleus and environment for the purposes of identifying the prime landing site for the Philae lander **[20]**, but also to provide the first scientific description of the comet with which to underpin all subsequent science of the mission.

Philae's descent and landing on the comet **[21]**, initial targeted the Agilkia landing site, but finally ended up resting at the Abydos site. The ~ 7 hour descent and subsequent multiple touchdowns on the comet surface facilitated unprecedented dual –spacecraft magnetic field measurements of the local environment and indicated the nucleus to have no intrinsic magnetic field on length scales > 1 m **[22]**. A single 1 mm dust particle was detected ~ 2.4 km from the surface, with comparisons to laboratory experiments suggesting a bulk density of 250 kg/m$^3$, likely being a porous conglomerate [23]. Subsequent detections were likely hampered by detector obscuration and operation times, although upper limits on millimetric particle flux (1.6 x 10$^9$ m$^{-2}$sr$^{-1}$) and volume density 10$^{-11}$ m$^{-3}$ – 10$^{-12}$ m$^{-3}$ on and near the surface have been provided [24]. Philae's approach revealed the surface of the comet to be photometrically uniform, with average brightness of the surface notably constant, appearing granular at 1 cm resolution, covered by regolith composed of debris and blocks from cm's to 5 m in scale [25]. The average normal albedo of Agilkia is 6.7% [26], slightly higher than the overall albedo of the comet [27] and was considered to have a ~20 cm thick granular soft surface layer (compressive strength ~ 1 kPa), with a much harder, sintered sub surface [28] similar to the surface of the Abydos region [29], which was found to have compressive strength of ~ 2 MPa [30]. The journey from Agilkia to Abydos facilitated observations of the near nucleus coma, although more likely of the surface material perturbed by the initial impact, revealing CHO-bearing organic compounds [31] and volatile ratios and further organics revealed at Abydos [32] [33]. From the Abydos site, internally the head lobe was found to be structurally homogeneous on scales of tens of metres, with a porosity of 75-85% and a dust to ice ratio of 0.4-2.6 [34] with suggestions of changes of the dielectric properties with depth [35], which may be related to changing porosity [36]. Abydos was shown to have a highly complex terrain of fractured and varying scale, with bright cm and mm scale features potentially indicating ice [37], and albedos varying from 3-5% compared to 5,8-6.7% for the area surrounding Abydos [38]. A more comprehensive review of the Philae results can be found in [39] and [40].

Based on the initial shape models of the comet [41] the smaller lobe or "head" of the "duck" was shown to be 2.6x2.3x1.8km and the larger lobe or "body" of the "duck" was 4.1x3.3x1.8 km, with most recent estimates on volume of 18.7±0.3 km$^3$, mass of 9,982±3 x10$^9$ kg and density of 533 ± 6 kg/m$^3$ [42]. The comet was revealed be morphologically highly diverse, and a number of terrains were identified, classified according to appearance: brittle material; dust-covered terrain; large scale depression structures; smooth terrain and consolidated exposed surfaces. These regions were assigned names of Egyptian deities, female for the upper lobe and male for the lower lobe [43][44]. Smooth thin deposits of dust found in the northern hemisphere were shown to be a result of "airfall" of non-escaping large particles emitted from the neck region of the comet. Dust transport in this manner was also shown to be capable of driving surface features such as aeolian ripples and ventifacts [45] and the 'splashing' discussed by [25]. Meter-scale fracturing was revealed to be a ubiquitous feature of the more consolidated regions, where thermal insolation weathering was considered responsible for these features, with potential to aid surface evolution and erosion [46]. The mechanical properties of the surface in a number of different regions were constrained by comparing gravitational slopes and surface morphology [47], where low slope (0-20°) terrains contained mainly fine material and few large isolated boulders (> 10 m), intermediate slope terrains (20°-45°) associated with fallen consolidated material, debris fields with numerous intermediate size boulders (< 1m – 10m) and high-slope terrain (45-90°) being cliff regions with exposed consolidated material with no boulders or fine material. Here consolidated is used to refer to areas that appear rocky in appearance and are cohesive enough to display lineaments and fractures.
Overhang compressive strength ranged from 3-15 Pa (upper limit 150 Pa), a 4-30 Pa shear strength range for boulders and fine material and 30-150 Pa for the compressive strength range of overhangs (with an upper limit of 1500 Pa). Such tensile strengths favour the formation of comets by the accretion of pebbles at low velocity. However, these values are significantly different to the compressive strengths > 2 MPa at the Agilkia site [48], based on measurements made by the Multipurpose Sensors for Sub-Surface Science (MUPUS). These were associated with a sintered dust ice subsurface layer comparable to laboratory experiments which have shown the formation of hard subsurface layers via sublimation/re-deposition cycles. Such processing was also discussed in the context of exposed ice patches observed at various locations on the surface [49], highlighting the diversity of surface geology. The Imhotep region (on the "belly" section of the "body" of the "duck") is the focus of great interest due to its location on the equator and hence experiencing illumination throughout the orbit around the sun. [50] provided a first analysis of the geomorphology of this region, in particular focusing on the basin like regions, suggesting formation by sub-surface void and subsequent collapse, as well as round elevated structures formed by exhausted outgassing vents being filled and surrounding terrain being eroded over the period of the orbit (and hence activity cycle). Boulder size distribution analysis of the northern illuminated portion of the comet revealed 3546 boulders > 7 m with a power law index of -3.6 +0.2/-0.3 with only slight differences in the cumulative size- frequency distribution between the small (-4.0 +0.3/-0.2) and main lobe (-3.5 +0.2/-0.3), but large differences between when compared the neck region (-2.2 +0.2/-0.2). Similar size-frequency distributions were reported for similar geomorphological settings, some on opposite sides of the comet, suggesting similar processes are active in these regions [51]. Boulder distribution and formation mechanisms may include activity outbursts and gravitational collapse, impacts (although there is evidence for only 1 impact crater on the surface [43]), boulder lifting and fragmentation and sublimation. Power law-indices for terrestrial analogues range from -2 to -3, although volcanic ash and pumice at -3.54 provides a better match to C67P, perhaps linked to the formation via fragmentation due to subsurface volatile overpressure and release. Along with boulders, pit structures are a common feature on the nucleus surface of 67P and other comets **[52]**. **[53]** suggest these features (primarily found in the Seth and Ma'at region) are formed by subsurface sublimation and subsequent cavity and collapse, somewhat similar to a sinkhole on Earth, with the regions growing via sublimation processes and their size and distribution indicating heterogeneity of the first few hundred meters of the surface. We note that this is in contrast to the structural homogeneity reported on the global scale i.e. [34 ][42] and will be addressed by subsequent studies examining the closest orbits. **[54]** also support the sinkhole hypothesis, but consider a much more recent formation via clathrate destabilization and amorphous ice crystalisation than considered by **[52]**. Subsurface void evolution (primordial or newly formed via sublimation) and subsequent collapse was also discussed by **[55]** in terms of the various boulder field formation mechanisms in the Aswan site in the Seth region. This region is characterized by layered terrain with associated cliffs and talus deposits. These strata pervade other regions across the comet and were used to examine the source of the bi-lobed shape of the nucleus by studying their orientation with respect to gravity. **[56]** have shown that the overall ordering of these layers in the comet is such that the comet was formed by a gentle impact of two similar yet independently formed cometesimals, in the early stages of the solar system, as had previously been suggested by **[57]**. More details on the bulk properties of comet 67P in comparison to other comets can be found in **[58]**.

Early observations revealed two populations of dust: one up to 2 cm in size and outflowing from the comet and detected within about 20 km of the spacecraft and a bound population, with particles ranging in size from 4 cm to ~2 m, detected at distances > 130 km from the spacecraft **[59]**. The overall population collected by Rosetta at heliocentric distances > 3 AU was rather fluffy and devoid of volatiles, with the coating of dust covering a parts of the nucleus surface thought to be a result of build up since the previous perihelion and early activity was beginning to remove this layer **[60]**.

No evidence for satellites were observed for objects > 6 m within 20 km of the nucleus and none > 1 m between 20 and 110 km **[61]** roughly in agreement with the upper limits in [59]. However, four objects in the range 0.14 - 0.5 m were observed, three with elliptical orbits consistent with an orbiting cloud, although one of the objects could have originated from the surface shortly before observation **[62]**. Initial gas and dust emissions revealed a dust to gas ratio of 4 +- 2 for 3.7 – 3.4 AU **[59]** (6+-2 if only water is considered) and 3.8-6.5 between 4.5 and 2.9 AU **[17]** (with a dust loss rate evolution from 3.7-2.9 AU of 0.5 - 15 kgs$^{-1}$), which when combined with the other physical characteristics above, imply a rather porous nucleus (~72-74%), that is an icy dust ball rather than a dusty snowball, and one that is rather homogeneous down to 10- 100 m scales **[42][34]**. Dust activity increased with decreasing heliocentric distance by a factor 6 between 3.36 and 2.43 AU, and two distinct populations of dust became apparent: fluffy aggregates (0.2-2.5 mm, with densities ~<1 kg/m$^3$) and compact particles (80-800 micrometres, densities (1.9+-1.1) x 10$^3$ kg/m$^3$) **[63]**. The fluffy particles had no specific source location, being detected over a range of latitudes and longitudes. These particles are also considered to be susceptible to fragmentation effects of the spacecraft electrostatic environment **[64]**. Observations of 100eV/q-18keV/q negative particles, with lower energies (200-500 eV) from the comet direction and higher energies (1-20 keV) from the sunward direction, have been presented as the first measurement of energetic charged sub-micron dust or ice grains (nanograins) in a cometary environment **[65][66]**. Solar radiation pressure also perturbs the micron and submicron population, with a factor three higher flux from the sunward direction compared to the comet nucleus. The compact particles are correlated with lower phase angle observations (between 30° and 40°) and with the neck region of the comet, with velocities of 2.5 +- 0.8 ms$^{-1}$ at 10 km from comet centre and 4.3 +- 0.9 ms$^{-1}$ at 30 km **[64]**. Photometry of coma dust grains show agreement with surface values, although some differences in composition from the surface could potentially indicate the presence of hydrated minerals **[67]**. A more detailed characterization of the dust environment was made following the collection of over 10, 000 particles of scale of a few 10 micrometres to several 100 micrometres during the period August 11, 2014 – April 3, 2015 ( 3.57 – 1.95 AU) **[68]**. In this case the particles are categorizes as clusters (~ fluffy particles from **[59]**) and compact, where there are three types of clusters: shattered, glued clusters and rubble piles. These cluster subcategories are based on the size distribution, spatial relationship between components and existence, or not, of a connecting matrix. A majority of these types of particles are considered to have originated from parent particles > 1mm in size, but have fragmented during entry into the instrument. The compact particles constitute 15% of the > 100 micrometre population and temporal correlations from sample collection indicate that they may have a common parent particle to clusters, contrary to suggestions from **[59]** that the clustered/fluffy particles and compact particles constitute distinct populations. At smaller heliocentric distances, the compact components correspond to > 30% of the fragments of parent particles (50-100 micrometre) suggesting an evolution from fluffy to parent particles with more compacted sub-particles. This is consistent with suggestions by **[60]** that activity increase would remove the initial fluffier dust layers. Analysis of the a broad sample of dust particles (585 particles of size > 14 micrometres) obtained between 3.6-3.1AU revealed a size distribution index of -3.1 **[69]**, with no clear evidence of organic matter and composition and morphology similar to interplanetary dust particles. Models of the dust trail of the comet indicate a dominance of mm sized particles **[70]**, the lower end of the size distribution feasibly elevated by models of gas driven activity at the surface **[71]**, which have not been able to explain the existence of smaller grains in the coma, suggesting they are a product of spin disintegration of larger grains. However, observations of rotating coma grains suggest this may not be the case **[72]**. Further discussion of the dust characteristics can be found in **[73]** and **[74].**

The nucleus surface was revealed to be an organic rich and very dark, a highly dehydrated surface with an upper limit on surface water ice abundance of only ~1% during the comet approach phase, with an albedo of 0.06 ±0.003 in the visible and infrared band **[75]** and down to 0.041-0.054 in ultraviolet wavelengths **[76][77]**. The nucleus exhibited phase reddening in the visible and infrared **[27] [78]**, with a variability of spectral slope across the nucleus, yet with no overall discernable variability that would distinguish either of the two lobes. Strong opposition effects were observed and spectral slope values were anti-correlated with reflectance, with the Hapi region in the neck being brightest and bluest. Analysis of the colour of the entire surface (variegation analysis) with respect to activity suggests that active regions have bluer spectra than the overall surface **[79]**. The inference of surface ice patches **[41]** was followed up more rigorously by **[49]** with these regions having a significantly bluer spectrum when compared to the surrounding terrain and being predominantly located in regions of low insolation, near the foot of cliff regions. These features were shown to persist over a period of months, and those located in the Imhotep region, on the 'belly' of the duck, were identified as water ice **[80]** with a bimodal distribution of ice grains in the micrometre and mm size range. The larger grain sizes were associated with sintering or sublimation driven grain growth by subsurface recondensation, whereas the finer grains were associated with re-condensation in ice-free layers. These micrometre grains were also seen in the Hapi region in the "neck" of the comet **[81]**, exhibiting a diurnal cycle following local illumination conditions, where subsurface sublimation continues for a short while as the comet surface rotates into the night, replenishing the surface with water ice. Temperature depth profiles inferred a thermal inertia range ~10-50 $JK^{-1}m^{-2}s^{-0.5}$ indicating large differences between surface and sub-surface temperatures, consistent with a thermally insulating powdery surface **[82]**. This value was better constrained by **[83]** to 10-30 $JK^{-1}m^{-2}s^{-0.5}$ using a simple homogenous model, with indications of inconsistencies potentially driven by vertical structure in the physical properties of the upper few cm of the surface. Observations of the un-illuminated southern regions of the comet agree with these overall values **[84]** with indications of ices within the first few 10's cm.

The first detection of water was made by MIRO on 6 June 2014, when the comet was 3.92 AU from the Sun **[82]**. Diurnal varations were seen, a gas expansion velocity of about 400 m/s was measured, and the total amount of water coming from the comet averaged about $10^{25}$ molecules/sec. These measurements are significant because they help characterize cometary water activity at large heliocentric distances and the distribution of near-surface water ice while far from the Sun. The water production rate during this period was consistent with surface observations as only ~1% surface ice was required to support observed production, which grew from 0.3 kg/s in June 2014 to 1.2 kg/s in August 2014 **[82]**. Diurnally the production rate varied by a factor of 2 and spatially the production rate varied by a factor 30, from $0.1 \times 10^{25}$ molecules $s^{-1}$ $sr^{-1}$ to $3.0 \times 10^{25}$ molecules $s^{-1}$ $sr^{-1}$ **[85]**, with highest column densities observed in the neck region **[86]**. $CO_2$ outgassing was observed in illuminated and un-illuminated regions, suggesting a subsurface source **[87]**, with $CO_2/H_2O$ from the neck region ~2%. Overall the coma showed strong heterogeneity in $H_2O$, CO and $CO_2$ roughly in agreement with a $1/R^2$ dependence in coma density, where R is distance from the comet's centre **[88]** and other volatiles such as $C_2H_6$, $CH_3OH$ and HCN **[89]** with variations strongly tied to the rotation period and comet latitude but also linked to differential sublimation at depth and potentially different phases of ice within the nucleus. The Imhotep region on the "body" or larger lobe exhibiting deep minima in $H_2O$ but localized maxima in $CO_2$ and largest abundances of $H_2O$ were observed in the "neck" region of the nucleus. Indeed, insolation driven models of the neutral coma with enhanced northern latitude activity provided better fits to insitu density measurements **[90]**. In the case of the dust and gas coma, outgassing enhancements in the Hapi neck region and neighbouring Hathor region, with dust size distributions from **[59]**, were required to better fit observations **[91]**. The broad volatile inventory of the comet at large heliocentric distances (~3.1 AU) **[92]** exhibits winter and summer hemispheric differences and significant abundances of CO and $CO_2$ when compared to other Jupiter-family comets (JFC). For example, at 67P CO ranges from 2.7% and 20% relative to water, which is comparable to 13% at 1P/Halley but much higher than 0.15-1% at 103P/ Hartley. $CO_2$ varies between 2.5% and 80% compared to 2-4% at 1P/Halley and 7-20% at 103P/ Hartley. These values for 67P are from outside 3 AU, whereas most other values are for comets at smaller heliocentric distances with higher water production rates. The deuterium to hydrogen (D/H) ratio of 67P/Churyumov-Gerasimenko was found to be $(5.3 +- 0.7) \times 10^{-4}$ **[93]**, much higher than previous observations of JFCs such as 103P/Hartley **[94]**, suggesting a broad and diverse origin of JFCs and further complicating the issue of the origin of terrestrial water from cometary impacts. Coupled with an argon to $H_2O$ ratio of $(0.1$ to $2.3) \times 10^{-5}$ **[95]** suggests JFCs are not a major volatile source for the Earth. The first detection of molecular nitrogen ($N_2$) **[96]**, where $N_2/CO = (5.70 +- 0.66) \times 10^{-3}$, and molecular oxygen ($O_2$) **[97]**, with a mean abundance of $O_2/H_2O$ of 3.80 +- 0.85% coupled to the high D/H suggest 67P/Churyumov-Gerasimenko was formed at very low temperatures. The observed abundances and heterogeneity have led to discussions regarding the nature of the ice within the comet **[98] [99]**. The 10 km orbits prior to lander deployment also facilitated the first detection of the amino acid glycine and also phosphorus [100], adding to the top of the list of the veritable volatile zoo detected by Rosetta. A broader discussion on ice and coma chemistry can be found in [101] and [102].

The interaction of the solar wind with this outflowing coma was registered locally by Rosetta in August 2014 when the comet was ~3.6 AU from the Sun, with the observation of accelerated water ions [103] as well as the breakdown of volatiles through a combination of photo-ionization and photoelectron impact disassociation [104,105]. Initial interactions of the comet with the solar wind were found to be highly turbulent and stronger than expected with the observation of surathermal electrons consistent with much higher activity comets (outgassing at 69P/CG is a factor 100 less than 1P/Halley), yet no upstream bow shock was detected [106]. Access to near nucleus altitude (< 30km) of the low activity comet facilitated the study of ion-neutral chemistry [107], which revealed inconsistencies with model predictions related to the heterogeneous coma as well as observation of H ions formed by double charge exchange of solar wind with molecules in the coma [108]. These close distance also facilitated the observation of the direct interaction of the solar wind with the nucleus resulting in sputtering of dust [109] revealing the comet refractories to have similar Na abundances to carbonaceous chondrites, a depletion in Ca and an excess of K. We note that these Na values are not consistent with [60] who report preliminary values of high Na abundances (> IDPs and chondrites). There are known contamination issues with Na [110] but this abundance is reported to persist throughout the mission [111]. The first measurement of pick-up ions was reported at ~3.5 AU [112]. From 3.6 – 2.7 AU a persistent "comet song" was detected, in the form of low-frequency compressional magnetic oscillations ~40 mHz, where classical pickup=ion driven instabilities were unable to explain the observations and instead a cross-field current instability was suggested as a possible source [113]. Such waves were also investigated using Philae and Rosetta during the Philae landing, revealing 278 km wavelength waves to have a phase velocity of ~ 6 km/s [114]. The flux of accelerated water ions (with energy > 120 eV) increased with approach to the Sun (from 3.6 – 2.0 AU [115]) with significant solar wind deflection observed, up to 45° from the anti sunward direction (in some cases > 50° [116]), with changes in the deflection associated to the changes to the orthogonal component of the Interplanetary Magnetic Field (IMF) [117]. The spatial distribution of the low energy plasma (1-10's eV) in the vicinity of the nucleus was highly structured (with strong peaks in the northern hemisphere above the neck), indicating strongly that the main source was ionization of the neutral coma. The electron density exhibited a 1/R fall of with distance, as expected from the ionization of a neutral expanding gas, although this could also be a combination of the effects of solar wind electric field and transport. No boundaries between solar wind and the cometary coma or separate plasma region signatures were observed beyond 3.1 AU [118]. The spacecraft potential remained negative within 50 km from the nucleus [119] and the plasma density increased, most significantly over the southern hemisphere, consistent with insolation increase and seasonal effects on the nucleus. Further discussion of the plasma phenomena investigated thus far by Rosetta can be found in [120].

Ground based observation from previous apparitions had indicated a non-isotropic coma [121], which were clearly apparent once Rosetta began to fully resolve the coma. Jet like features were resolvable from numerous regions on the surface, including Hapi, Hathor, Anuket and Aten, [122]. For the case of the jets emanating from the Hapi region, overall activity persists over many comet rotations, but the morphology of the jets evolves on times scales of single rotations to several days [123]. These jets were further scrutinized by [124] and linked to pit regions [53] and fractured cliffs [e.g. 55] on the surface; they are well correlated with solar illumination and hence volatile sources within the diurnal thermal skin depth of the surface. Simple insolation driven sublimation on the complex shape of the nucleus [125] is sufficient enough to enhance energy input in concave regions, such as the Hapi region and Seth pit regions through self-illumination [126] as well as cliff regions, which lead to surface planation following cliff wall erosion.

## Development of cometary activity, comparing with pre-perihelion and the extension

On 28th March 2015, the spacecraft encountered navigation issues due to the dust environment around the comet [13], eventually resulting in a safe mode on the spacecraft, after which the spacecraft retreated to a lower density region of the comet coma at ~400 km. From this period the overall trajectory plan of the mission was altered to follow the "simple" high level trajectory goal of getting as close to the nucleus as possible, while monitoring the capability of successful navigation in the comet dust environment. This included specific excursions to greater distances from the comet in September 2015 (~1500 km in the terminator/dayside direction) and March 2016 (~1000 km in the tailward direction) to investigate the outer coma. This evolution is clearly seen in Figure 1 and Figure 2 showing the overall trend of the spacecraft being at larger distances due to enhanced cometary activity (and hence more a intense dust environment) around centred at few weeks after perihelion on 12 August 2015, in line with previous observations of activity [7][127], as well as the excursions and fly bys.

A key driver of activity over the orbit of 67P/CG is its spin obliquity and nucleus shape **[128]**, where the southern latitudes of the comet only receive sunlight for around 10 months out of its 6.55 year orbit. In the case of the current apparition, this corresponds to May 2015 – March 2016 and following a number of months mapping the southern hemisphere, 7 regions were added to the 19 previously identified to complete the full region classification of comet 67P **[129]**. The southern hemisphere appears quite different to the northern hemisphere and is predominantly characterised by consolidated terrains, devoid of wide scale smooth terrains, dust coatings and depressions, somewhat contrary to the expectations of **[126]**. Comparison of the boulder population of the southern to the northern hemisphere recovered a similar power law index of -3.6 +/0.2 for diameter ranges 5-35m, suggesting common evolutionary processes. However, the southern hemisphere has a higher number of boulders (factor 3) then the northern hemisphere, suggesting more intense thermal fracturing and activity **[130]**. This southern summer season also revealed asymmetries in the major volatile abundance between the northern and southern hemisphere **[131][132][133]** with $CO_2$ more abundant in the South. Continued observations showed that the dust-to-water ratio persisted close to 6 for the entire inbound passage of the comet (3.6 AU to perihelion), with an evolution in the < 1 mm dust size distribution **[134]**. Further observations of Glycine during fly-bys at different radial distances suggest a distributed source of glycine associated with dust **[100]** and compositional analysis of dust particles collected in May 2015 show evidence of calcium–aluminum-rich inclusions (CAIs), previously found in Stardust samples **[135]** and high molecular weight organic matter in dust grains collected in May and October 2015 **[136]** which was not observed in particles obtained earlier **[69]**.

Although there were indications of surface changes in terms of mass wasting near cliffs and granular flows near pits **[124]**, the first major temporal changes appeared the smooth areas of the Imhotep region in the form of rounded features growing in size at **a rate of 5.6-8.1 x $10^{-8}$ $ms^{-1}$ during the period 24 May 2015- 11 July 2015 [137].** These regions are characterized by bluer spectra, indicating ice exposure or possibly hydrated minerals **[67][79]**, although comparisons with laboratory samples suggest this is not the case **[138]**. In the case of ice, erosion rates are 1-2 orders of magnitude lower than that observed, with speculation of this enhanced erosion being driven by low tensile strength of the surface matrix, or clathrate destabilization/amorphous ice crystallization, although observation of Sulphur bearing species in the comet have been used to argue against the existence of clathrates **[139]**. Shortly before equinox, sunset jets were observed in that Ma'at region **[140],** where jet activity persisted after sunset for ~ 1 hour. These jets were considered to be dayside driven activity, in dusty terrains, sustained by sub-surface thermal lag facilitating continued water ice sublimation, where the uneven distribution of these jets potentially relating to subsurface inhomogeneities. This study also highlighted subtle morphological surface changes of so called "Honeycomb" features, small scale pitted features in the Ma'at region associated with overall dust activity. The approach to perihelion saw activity increasing, including the observation of a curved jet **[141]** formed of 0.1-1 mm dust particles emitted from close to the equator in the Nut, Serqet and Ma'at region, enhancing the spiral feature of the jet. Initial predictions of rotation rate evolution dependence on insolation driven torques show good agreement with observations **[126][142]** and the close orbits of the final phase of the mission should allow us to examine in great detail the internal structure of the comet and address its level of heterogeneity, in particular the differences between local and global characteristics **[143]**.

The larger comet-spacecraft distances flown during 2015 meant that the exploration of the diamagnetic cavity was considered not possible, as it was expected to only a few 10s km from the nucleus. However, **[144]** report the detection of a cavity much larger than predicted by simulations **[145]**, suggesting a combination of low magnetic pressure in the solar wind and propagating instabilities on the cavity boundary facilitating the observations. Solar wind – cometary plasma interaction regions and boundaries were observed from mid-April 2015 onwards, with no solar wind signal observed until after January 2016 (aside from a short period during the passage of a CME coincident with the September 2015 excursion). Regions were characterized relative to one another behavior by either enhancement or reduction of various parameters such as electron density, water group ions energies and magnetic field. Located outside of the magnetic cavity boundary **[144]** and shown to be a permanent feature of the solar wind interaction at 67P, the ion velocity drop and electron density enhancements associated with the boundary resembles the ion pileup region observed at 1P/Halley **[146]**. These boundaries are related to the production rate of the comet and likely to be significantly affected by the outburst events that characterized observations around perihelion **[147]**. These transient releases of gas and dust identified in visible wavelengths occurred ~ 2.4 comet rotations and in total, 34 were observed, occurring in the early morning or shortly after local noon. This separation was explained to be driven by different underlying processes – morning outbursts being driven by thermal stresses of change of temperature and afternoon events driven by the thermal wave reaching volatile located deeper in the sub-surface. However, some events do not fit these mechanisms, instead being related to cliff collapse. Such a mechanism was suggested to be the driver of an event with broad instrument coverage on 19 February 2016 **[148]**, including visible and far-ultraviolet observations. A set of far-ultraviolet observations around perihelion of sporadic gas outbursts had no coincident visible signatures **[149]** demonstrating the importance of multi-wavelength and multi-instrument study of these phenomena over all heliocentric distances, with the latest occurring (at the time of writing) on 3 July 2016.

As indicated, these initial results are already being used to address questions about the origin of the comet and its place in the solar system: The implications of its bi-lobed structure **[56][57]** and its possible future **[150]**; Whether the comet originated from fragments from parent bodies in a collisional evolution **[151][57]** or whether it was a primordial rubble pile **[152]**; Putting these initial observations into context with other comets, e.g. **[153][50]**; Examining the connection between these bodies and the inner solar system **[154]** and Earth **[93];** Addressing the general questions of complex disk chemistry in primordial nebula **[155]**.

# Conclusion

The final phase of the Rosetta mission runs from the beginning of August 2016 to the end of September 2016, with the orbit plane tilted ~ 20° from terminator, and beginning with a pericentre on the dayside of 8 km, and apocentre on the night side of 13 km, the orbit will gradually be brought closer to the surface until on 30 September, following a phasing manoeuvre, Rosetta will embark on a trajectory towards the surface of the comet, with nominal touchdown at 30 September 2016 10:30:00 UTC. This final descent will entail acquisition of data from a number of instruments as late as possible, enabling observations of features at unprecedented resolution and offering new views of coma-surface interactions. Although operations will cease at impact, activity focusing on the archiving of the data will continue, to ensure all data from the mission is in a scientifically useable state for the future. This is a vital task, exemplified by recent Giotto data being re-analysed in light of Rosetta results **[156]**. As such, it will likely mark an increase in science activity, at the very least as the instrument teams will have the burden of operations lifted.

The Royal Society discussion meeting on "Cometary Science after Rosetta" provided an excellent forum for its subject matter. This paper is derived from a presentation designed to give a snap shot of the Rosetta mission science so far, particularly focusing on the orbiter, with the lander discussed more broadly elsewhere in this issue **[39]** as well as the significant ground and near –Earth campaign **[157]**. When considering what Rosetta set out to achieve, referring to table 1, it is the authors' belief that we have made great inroads to all aspects of the topics that the mission was designed to address, but we are by no means finished and there is plenty of science to come as correlative analyses among instruments becomes more prevalent and more sophisticated modeling is accomplished. Although close to the end of the operations of the mission at the time of writing, the published science output of the mission is just starting to focus on the period around perihelion and just beyond, as perhaps exemplified by the length of the sections above. So this current paper really is the story so far.

It is certain that the Rosetta mission as a whole, that being the Rosetta orbiter and the Philae lander **[39]**, has made a significant impact on solar system science **[158]** and in particular on cometary science **[157][10][160][161]**. It is too early to say how significant however.

**Acknowledgments**
MGGTT would like to thank the Royal Society for hosting the science discussion meeting " Cometary Science after Rosetta, and also give special thanks to his co-organisers Dr. Matthew Knight, Professor Alan Fitzsimmons and in particular Dr. Geraint Jones, whose persistent efforts and energy made this possible.

Rosetta is an ESA mission with contributions from its member states and NASA. Rosetta's Philae lander is provided by a consortium led by DLR, MPS, CNES and ASI. We thank all elements of the Rosetta project and instrument teams, including public outreach and communications for their tireless efforts in making the mission an astounding success. A special acknowledgement is extended to all our families without whose support, this could not have been achieved. MGGTT would additionally like to thank A. Milas for being metal \m/.

Please acknowledge anyone who contributed to the study but did not meet the authorship criteria.

**Funding Statement**
Work performed by BJB and MC was done at the Jet Propulsion Laboratory under contract to the National Aeronautics and Space Administration.

**Data Accessibility**
NAVCAM data presented are available through the PSA archive of ESA and the PDS archive of NASA. PSA is accessible via: http://www.cosmos.esa.int/web/psa/rosetta.

**Competing Interests**
*'We have no competing interests.'*

**Authors' Contributions**
MGGTT was the primary author. BB and MC all contributed to the drafting and concept of this paper, as did NA who also aided with the construction and provision of figures. All authors approve the final submitted version.

…

13118. Edberg, N.J.T. et al., Spatial distribution of low-energy plasma around comet 67P/CG from Rosetta measurements, *GRL*, 42, 4263-4269, doi:10.1002/2015GL064233
119. Odelstad, E. et al., Evolution of the plasma environment of comet 67P from spacecraft potential measurements by the Rosetta Langmuir probe instruments, *GRL*, 42, 10,126-10,134, doi:10.1002/2015GL066599
120. Glassmeier, K.-H. et al., 2016, Interaction of comets with the solar wind: a Rosetta perspective, *Phil.Trans.R.Roc.A.*, This issue
121. Lara, L.M. et al., 2011, 67P/Churyumov-Gerasimenko activity evolution during its last perihelion before the Rosetta encounter, *A&A*, **525,** A36, doi:10.1051/0004-6361/201015515
122. Lara, L.M. et al., 2015, Large-Scale dust jets in the coma of 67P/Churyumov-Gerasimenko as seen by the OSIRIS instrument onboard Rosetta, *A&A*, **583,** A9, doi:10.1051/0004-6361/201526103
123. Lin, Z.-Y. et al., 2015, Morphology and dynamics of the jets of comet 67P/Churyumov-Gerasimenko Early-phase development, , *A&A*, **583,** A11, doi:10.1051/0004-6361/201525961
124. Vincent, J.-B. et al., 2016, Are fractured cliffs the source of cometary dust jet? Insights from OSIRIS/Rosetta at 67P/Churyumov-Gerasimneko, *A&A*, **587,** A14 doi:10.1051/0004-6361/201527159
125. Preusker, F. et al., 2015, Shape model, reference system definition , and cartographic mapping standards for comet 67P/ Churyumov-Gerasimnko – Stereo-photogrammetric analysis of Rosetta/OSIRIS image data, *A&A*, **583,** A133 doi:10.1051/0004-6361/201526349
124. Keller, H.U. et al., 2015, Insolation, erosion, and morphology of comet 67P/ Churyumov-Gerasimenko, , *A&A*, **583,** A34 doi:10.1051/0004-6361/201525964
126. Bertaux, J.-L. et al., 2014, The water production rate of Rosetta target Comet 67P/Churyumov-Gerasimenko near perihelion in 1996, 2002 and 2009 from Lyman $\alpha$ observations with SWAN/SOHO, *Planetary and Space Science*, 91, 14-19, http://dx.doi.org/10.1016/j.pss.2013.11.006
128. Fulle, M. et al., 2016, Unexpected and significant findings in comet 67P/ Churyumov-Gerasimenko: an interdisciplinary view, *MNRAS*, 462, Suppl_1:S2-S8.
129. El-Maarry, M.R. et al., Regional surface morphology of comet 67P from Rosetta/OSIRIS images: The southern hemisphere, *A&A*, **593,** A100, doi:10.1051/0004-6361/201628634
130. Pajola, M. et al., The southern hemisphere of 67P/ Churyumov-Gerasimenko: Analysis of the pre-perihelion size-frequency distribution of boulders >= 7m, *A&A*, **592,** A36, doi:10.1051/0004-6361/201628887
131. Mall, U. et al., 2016, High-time resolution in situ investigationof major cometary volatiles around 67P/C-G at 3.1-2.3 AU measured with ROSINA-RTOF, *APJ*, 819:126, doi:10.3847/0004-637X/819/2/126
132. Migliorini, A. et al., 2016, Water and carbon dioxide distribution in the 67P/Churyumov-Gerasimenko coma from VIRTIS-M infrared observation, *A&A*, **589,**A45, doi:10.1051/0004-6361/201527661
133. Fougere, N. et al., 2016, Three-dimensional direct simulation Monte-Carlo modeling of the coma of comet 67P/Churyumov-Gerasimenko observed by the VIRTIS and ROSINA instruments on board Rosetta coma, *A&A*, **588,**A134, doi:10.1051/0004-6361/201527889
134. Fulle, M. et al., 2016, Evolution of the dust size distribution of comet 67P/Churyumov-Gerasimenko from 2.2 au to perihelion, *APJ*, **821:19** (doi:10.3947/0004-637X/821/1/19)
135. Paquette, J.A. et al., 2016, Searching for calcium-aluminium-rich inclusions in cometary particles with Rosetta/COSIMA, *Meteoritics & Planetary Science*, 51, Nr. 7 , 1340-1352, doi: 10.1111/maps.12669
136. Fray, N. et al., 2016, High-molecular weight organic matter in the particle of comet 67P/ Churyumov-Gerasimenko, *Nature*, 538,72-74, doi:10.1038/nature19320
137. Groussin, O. et al., Temporal morphological changed in the Imhotep region of comet 67P/ Churyumov-Gerasimenko, *A&A*, **583,** A36, doi:10.1051/0004-6361/201527020
138. Quirico, E. et al., 2016, Refractory and semi-volatile organics at the surface of comet 67PChuryumov-Gerasimenko: Insights from the VIRTIS/Rosetta imaging spectrometer, *Icarus*, 272, 32-47, doi:10.10/j.icarus.2016.02.028
139. Calmonte, U. et al., 2016, Sulphur-Bearing species in the coma of comet 67P/Churyumov-Gerasimenko, *MNRAS*, 462, Suppl_1:S253-S273.
140. Shi, X. et al., 2016, Sunset jets observed on comet 67P/Churyumov-Gerasimenko sustained by subsurface thermal lag, *A&A*, **586,** A7, doi:10.1051/0004-6361/201527123
141. Lin, Z.-Y. et al., 2016, Observations and analysis of a curved jet in the coma of comet 67P/Churyumov-Gerasimenko, *A&A*, **588,**L3, doi:10.1051/0004-6361/201527784
142. Bertaux, J.-L., 2015, Estimate of the erosion rate from $H_2O$ mass-loss measurements from SWAN/SOHO in previous perihelions of comet 67P/ Churyumov-Gerasimenko and connection with observed rotation rate variations, *A&A*, **583,** A38, doi:10.1051/0004-6361/201525992
143. Gutiérrez, P.J. et al., 2016, Possible interpretation of the precession of comet 67P/Churyumov-Gerasimenko, *A&A*, **590,** A46, doi:10.1051/0004-6361/201528029
144. Goetz, C. et al., 2016, First detection of a diamagnetic cavity at comet 67P/ Churyumov-Gerasimenko, 23, *A&A*, **588**,A24, doi:10.1051/0004-6361/201527728
145. Koenders, C. et al., 2015, Dynamical features and spatial structures of the plasma interaction region of 67P/ Churyumov-Gerasimenko and the solar wind, *Planetary and Space Science*, 105, 101-116, doi: 10.1016/j.pss.2014.11.014
*Phil. Trans. R. Soc. A.*

**Tables**

Table 1. Based on the Rosetta Science Management Plan [1].

| Rosetta Prime Scientific Objectives |
|---|
| <ul><li>Global characterisation of the nucleus, determination of dynamic properties, surface morphology and composition.</li><li>Chemical, mineralogical and isotropic compositions of volatiles and refractories</li><li>Physical properties and interrelation of volatiles and refractories in a cometary nucleus</li><li>Study the development of cometary activity and the processes in the surface layer of the nucleus and the inner coma (dust-gas interaction)</li><li>Origin of comets, relationship between cometary and interstellar material. Implications for the origin of the solar system</li><li>Global characterisation of the asteroid, determination of dynamic properties, surface morphology and composition.</li></ul> |

**Figure and table captions**

**Figures**

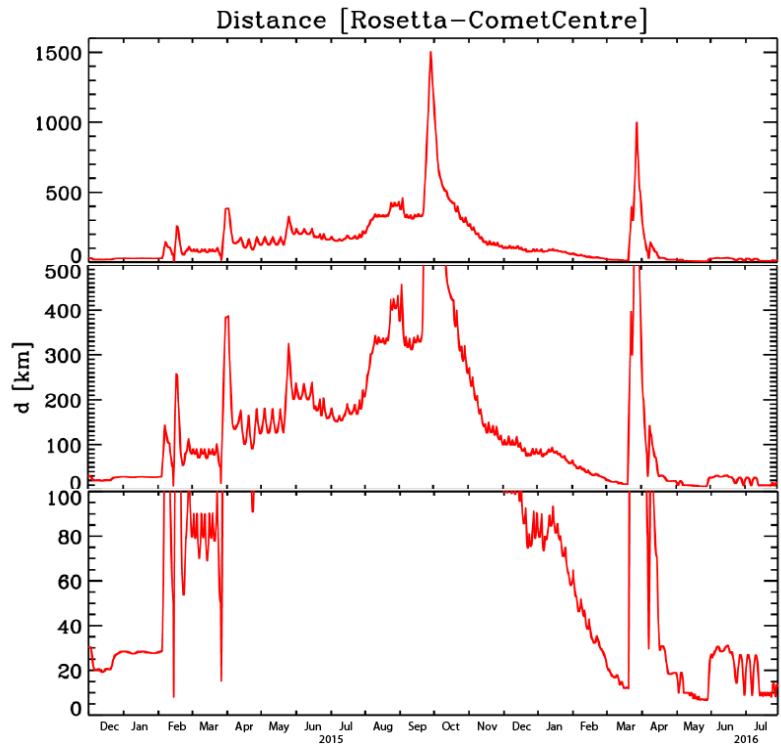

Figure 1. Evolution of comet- spacecraft distance. Lower panel shows orbit evolution in terminator plane over whole mission.

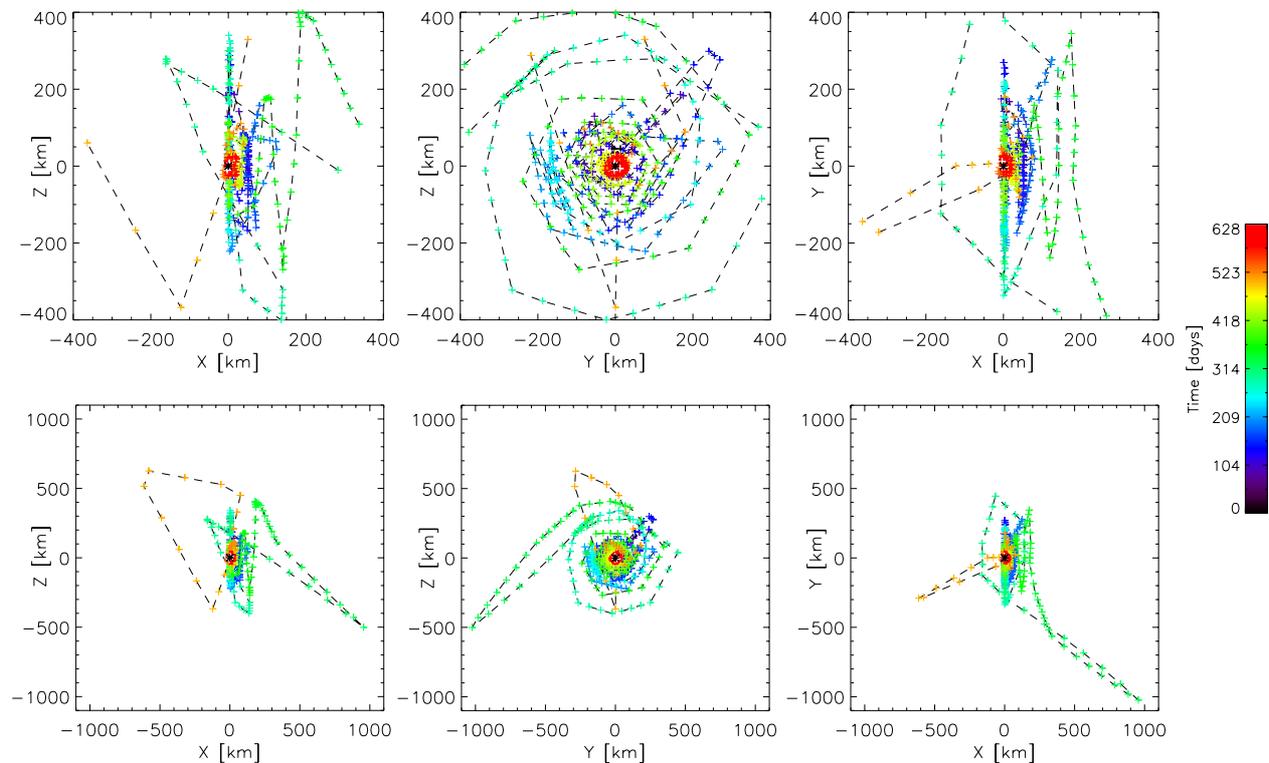

Figure 2. Rosetta orbiter trajectory plotted for each day from 1 January 2015 – 1 August 2016. X axis point from comet to the Sun, Y is along the projection of the comet heliocentric velocity vector and +Z completes the right handed frame. The colour coding indicates the elapsed time in days. The plot clearly indicates the amount of time spent at the terminator, and also the large excursions to the days side and night side of the comet, as well as fly by preparation.

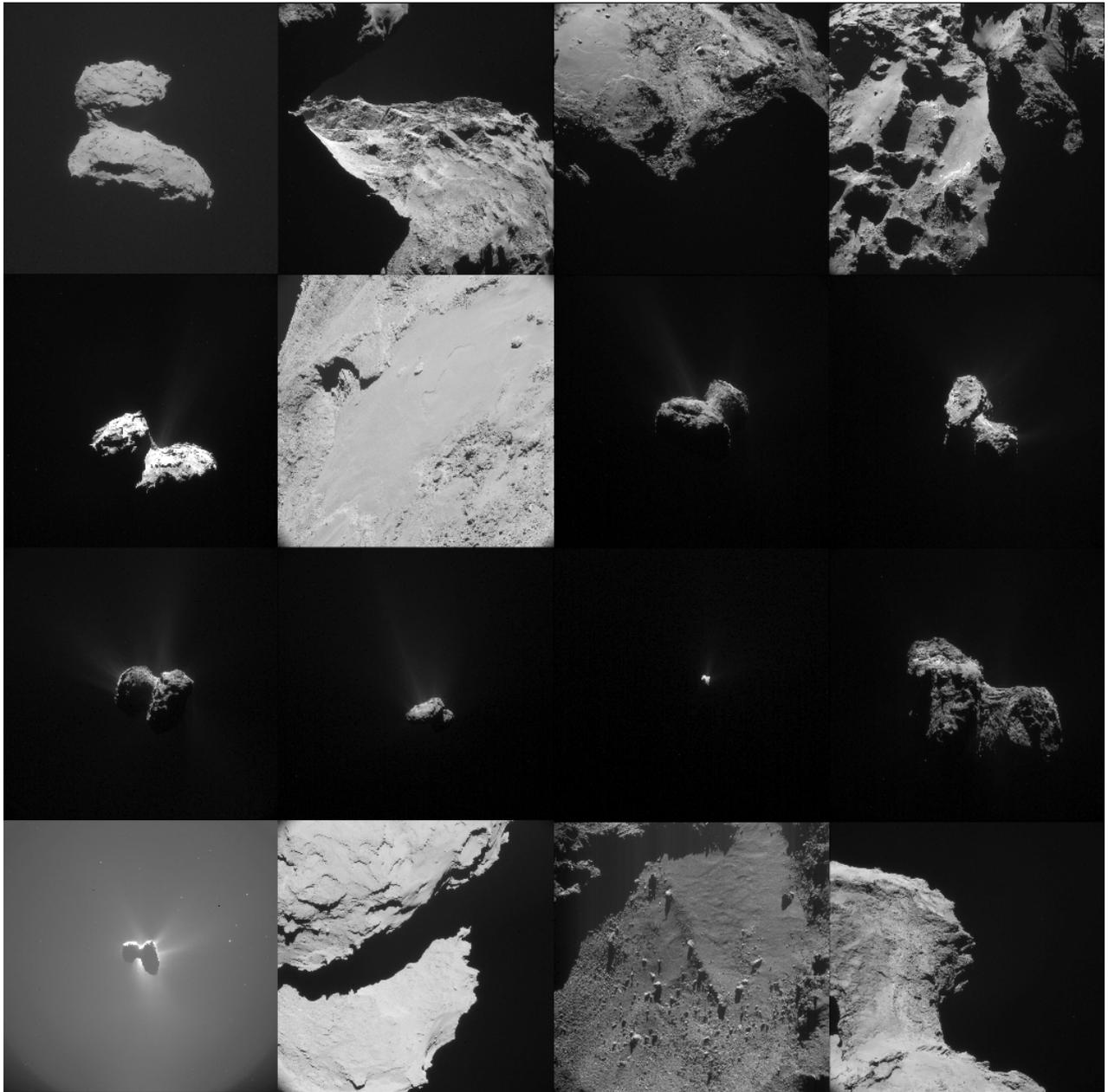

Figure 3. Collection of Rosetta navigation camera images from August 2014 - June 2016. Distances are from comet center. From top left moving clockwise: 7/8/2014 11:07:17 UT, from 84.920 km, 13/10/2014 06:22:55 UT from 18.183 km, 21/11/2014 19:47:54 UT from 31.076 km, 16/12/2014 05:29:34 UT from 20.525 km, 06/2/2015 14:47:55 UT from 124.016 km, 14/2/2015 14:19:43 UT from 10.641 km, 12/4/2015 20:25:02 UT from 149.123 km, 30/4/2015 00:27:01 UT from 155.435 km, 01/7/2015 14:51:35 UT from 159.588 km, 12/8/2015 14:51:35 UT from 332.379 km, 28/9/2015 21:54:23 UT from 1276.16 km, 2/1/2016 19:38:37 UT from 84.069 km, 27/3/2016 12:53:21 UT from 328.660 km, 09/4/2016 21:42:52 UT from 29.945 km,13 May 2016 23:04:22 UT from 9.959 km, 15 June 2016 15:03:30 UT from 29.262 km